\documentclass[aps,prMat,reprint,superscriptaddress,floatfix,longbibliography]{revtex4-1}
\usepackage{lipsum}

\usepackage{epsfig}
\usepackage{latexsym}
\usepackage{epic}
\usepackage{eepic}
\usepackage{psfrag}
\usepackage{rotating}
\usepackage{multirow,booktabs}
\usepackage{color}
\usepackage{cprotect} 
\usepackage{miller} 
\usepackage{xspace} 

\usepackage[export]{adjustbox}
\usepackage{graphicx}
 
\usepackage{algorithm}
\usepackage[]{algorithmicx,algpseudocode}

\usepackage[fleqn, leqno]{amsmath}
   \usepackage{amsfonts}   
   \usepackage{amssymb}    
\usepackage{amsthm}
\usepackage{stmaryrd} 
\usepackage{import}
\usepackage{breqn} 

\usepackage{array}

\usepackage{ulem}

\newcommand{\done}[1][]{{\color{green} \ensuremath{ \surd}\ifx\relax#1\relax\else(#1)\fi}} 
\newcommand{\prob}[1][]{{\color{red} { \bf!}\ifx\relax#1\relax\else(#1)\fi}} 
\newcommand{\fail}[1][]{{\color{red} \ensuremath{\lightning}\ifx\relax#1\relax\else(#1)\fi}} 

\newcommand{\ddr}{\ensuremath{{\rm d}{\bf r} }\xspace} 
\newcommand{\rr}{\ensuremath{{\bf r} }\xspace} 
 
\newcommand{\kk}{\ensuremath{{\bf k} }\xspace} 
\newcommand{\mm}{\ensuremath{{\bf m} }\xspace} 
 
\newcommand{\mmi}{\ensuremath{{ m_i} }\xspace} 

\newcommand{\BB}{\ensuremath{{\bf B} }\xspace} 
\newcommand{\BBext}{\ensuremath{{\bf B_{\rm ext}} }\xspace} 
\newcommand{\BBind}{\ensuremath{{\bf B_{\rm ind}} }\xspace}

\newcommand{\VA}{\ensuremath{{\bf A} }\xspace} 

\newcommand{\phii}[1][]{\ensuremath{\varphi}\xspace\ifx\relax#1\relax\else\ensuremath{\left(#1\right)}\xspace\fi}

\newcommand{\FF}[1][]{\ensuremath{\mathcal{F}}\xspace\ifx\relax#1\relax\else\ensuremath{\left[#1\right]}\xspace\fi}
\newcommand{\FFm}[1][]{\ensuremath{\mathcal{F_{\rm m}}}\xspace\ifx\relax#1\relax\else\ensuremath{\left[#1\right]}\xspace\fi}
\newcommand{\FFpfc}[1][]{\ensuremath{\mathcal{F_{\rm PFC}}}\xspace\ifx\relax#1\relax\else\ensuremath{\left[#1\right]}\xspace\fi}
\newcommand{\FFcoup}[1][]{\ensuremath{\mathcal{F_{\rm c}}}\xspace\ifx\relax#1\relax\else\ensuremath{\left[#1\right]}\xspace\fi}

\newcommand{\ff}[1][]{\ensuremath{f}\ifx\relax#1\relax\else\ensuremath{\left(#1\right)}\fi}

\newcommand{\ffpfc}[1][]{\ensuremath{\ff_{\rm PFC}}\ifx\relax#1\relax\else\ensuremath{\left(#1\right)}\fi}
\newcommand{\ffid}[1][]{\ensuremath{\ff_{\rm id}}\ifx\relax#1\relax\else\ensuremath{\left(#1\right)}\fi}
\newcommand{\ffex}[1][]{\ensuremath{\ff_{\rm ex}}\ifx\relax#1\relax\else\ensuremath{\left(#1\right)}\fi}
\newcommand{\ffm}[1][]{\ensuremath{\ff_{\rm m}}\ifx\relax#1\relax\else\ensuremath{\left(#1\right)}\fi}
\newcommand{\ffcoup}[1][]{\ensuremath{\ff_{\rm c}}\ifx\relax#1\relax\else\ensuremath{\left(#1\right)}\fi}

\newcommand{\NN}[1][]{\ensuremath{\mathcal{N}}\ifx\relax#1\relax\else\ensuremath{\left[{#1}\right]}\fi}

\newcommand{\LL}[1][]{\ensuremath{\mathcal{L}}\ifx\relax#1\relax\else\ensuremath{#1}\fi}

\newcommand{\np}[1][]{\ensuremath{#1}\ifx\relax#1\relax\else\ensuremath{^{n+1}}\fi}
\newcommand{\n}[1][]{\ensuremath{#1}\ifx\relax#1\relax\else\ensuremath{^{n}}\fi}

\newcommand \be {\begin{eqnarray}}
\newcommand \ee {\end{eqnarray}}

\makeatletter
\let\cat@comma@active\@empty
\makeatother
\begin{document}


\title{Magnetically induced/enhanced coarsening in thin films}


\author{R. Backofen}
\affiliation{Institute of Scientific Computing, Technische Universit\"at Dresden, 01062 Dresden, Germany}
\author{A. Voigt}
\affiliation{Institute of Scientific Computing, Technische Universit\"at Dresden, 01062 Dresden, Germany}
\affiliation{Dresden Center for Computational Materials Science (DCMS), 01062 Dresden, Germany}

\begin{abstract}
  External magnetic fields influence the microstructure of polycrystalline materials.
  We explore the influence of strong external magnetic fields on the long time scaling of grain size during coarsening in thin films with an extended phase-field-crystal model.  Additionally, the change of various geometrical and topological properties is studied.
  In a situation which leads to stagnation, an applied external magnetic field can induce further grain growth.
 The induced driving force due to the magnetic anisotropy defines the magnetic influence of the external magnetic field. Different scaling regimes are identified dependent on the magnetization. At the beginning, the scaling exponent increases with the strength of the magnetization. Later, when the texture becomes dominated by grains preferably aligned with the external magnetic field, the scaling exponent becomes independent of the strength of the magnetization or stagnation occurs. We discuss how the magnetic influence change the effect of retarding or pinning forces, which are known to influence the scaling exponent.  
 We further study the influence of the magnetic field on the grain size distribution (GSD), next neighbor distribution (NND) as well as grain shape and orientation.  If possible, we compare our predictions with experimental findings. 
\end{abstract}

\pacs{?}

\maketitle


\section{Introduction}

Grain boundaries in polycrystalline materials are of paramount importance to various fields of science and engineering. They have been intensively studied theoretically and experimentally over decades. Quantitative comparison of geometrical and topological properties between theory or simulation and experimental data are still unsatisfactory in general. Progress have been made for nanocrystalline thin metallic films. Geometric and topological characteristics of the grain structure can be shown to be universal and independent of many experimental conditions \cite{Barmaketal_PMS_2013}. A phase field crystal (PFC) model \cite{Elderetal_PRL_2002,Elderetal_PRE_2004}, which considers the essential atomic details but operates on diffusive time scales, was able to reproduce the universal grain size distribution and showed similar scaling properties and stagnation as in the experiments \cite{BBV14}. This is in contrast with more classical Mullins-like models, which only consider the evolution of the continuous grain boundary network \cite{Mullins_JAP_1956}. Theoretical predictions and simulations for this type of models lead to self-similar structures and coarsening laws for the average grain size of the form $t^\alpha$, with a scaling exponent $\alpha = 1$ in the original setting \cite{Mul98}.
These models have been extended by including retarding and pinning forces for grain boundary movement \cite{Barmaketal_SM_2006,HF10,Streitenbergeretal_AM_2014} and grain rotation \cite{Upmanyuetal_AM_2006,Tothetal_PRB_2015,Esedoglu_CMS_2016}. The modifications can explain the smaller scaling exponents in experiments and stagnation.
However, also these modifications are unable to reproduce the universal grain size distribution (GSD). A detailed comparison between these models with PFC simulations \cite{BBV14} and experiments in Ref.~\cite{Barmaketal_PMS_2013} can be found in Ref.~\cite{LaBoissoniereetal_M_2019}.
With the achieved agreement for various geometrical and topological properties it is now time to use the PFC model as a predictive tool to control grain growth in thin films under the influence of external fields.

External magnetic fields during processing influence grain growth and as such have been proposed as an additional degree of freedom to control the grain structure, see \cite{Riv13,GEV18} for reviews. The PFC model has been extended to include magnetic interactions in Refs.~\cite{FPK13,SSP15} and  was used in 
Ref.~\cite{BEV19} to explain the complex interactions between magnetic fields and solid-state matter transport.
An applied magnetic field influences the texture during coarsening due to the anisotropic magnetic properties of the single grains. Grains with their easy axis aligned to the external field are energetically preferred. They grow preferably at the expense of the other grains.  
The mobility of grain boundaries in this model is found to be anisotropic with respect to the applied magnetic field.
Magnetostriction is naturally included in the extended PFC model.
All these effects already change texture on small time scales. In this paper we analyze the long time scaling behavior and various geometrical and topological properties in grain growth under the influence of a strong external magnetic field. 

The paper is organized as follows: We first review the underlying PFC model, the physical setting and the considered numerical approach. Then we consider the coarsening regime and analyze various geometrical and topological measures. Finally, we discuss the results, explain our findings and draw conclusions. 

\section{Model and numerical approach}
The model in Refs.~\cite{FPK13,SSP15,BEV19} combines the rescaled number density \phii of the original PFC model \cite{Elderetal_PRL_2002,Elderetal_PRE_2004} with a mean field approximation for
the averaged magnetization \mm. The total energy,
\begin{eqnarray*}
  \FF[\phii,\mm] = \int \ffpfc[\phii] + \omega_B \ffm[\mm] + \omega_B \ffcoup[\phii,\mm] \ddr ,
\end{eqnarray*}
consists of contributions related to local ordering in the crystal, $\ffpfc[\phii]$, and the local magnetization, $ \ffm[\mm]$. The magnetic anisotropy is included by coupling density field and magnetization in the last term, $\ffcoup[\phii,\mm]$. $\omega_B$ is a 
parameter to control the influence of the magnetic energy. 

An extended PFC model (XPFC) is chosen in order to define the crystal structure \cite{GPR10},
\begin{eqnarray*}
\ffpfc[\phii]  &=&  \frac{1}{2} \! \phii[\rr]^2 \!-\! \frac{t}{6} \! \phii[\rr]^3 \!+\! \frac{v}{12} \! \phii[\rr]^4 \\
               &&\quad -\frac{1}{2}  \phii[\rr] \int C_2(\rr-\rr') \phii[\rr'] \ddr'  .
\end{eqnarray*}
The magnetization is governed by 
\begin{eqnarray*}
\ffm[\mm]  &=&  \frac{W_0^2}{2} \! \left( \nabla \! \cdot \! \mm\right)^2 
               \! + \! r_m  \frac{\mm^2}{2} \! + \! \gamma_m \frac{\mm^4}{4} \! - \! \mm \cdot \BB \! + \! \frac{\BB^2}{2} ,
\end{eqnarray*}
where $r_m$ and $\gamma_m$ control the magnitude of magnetization and $W_0^2$ the energy due to inhomogenities of magnetization \cite{FPK13,BEV19}. 
The magnetic anisotropy is modeled by coupling the density wave with magnetization \cite{FPK13},  
\begin{eqnarray*}
 \ffcoup[\phii,\mm]   &=&   - \omega_m \phii^2  \frac{\mm^2}{2} \! - \!  \sum_{j=1}^2 
\frac{\alpha_{2j}}{2 j} \left( \mm \cdot \nabla \phii \right) ^{2j}.
\end{eqnarray*}

In order to maximize
the anisotropy, as in \cite{BEV19}, a square ordering of the crystal is
preferred, which is realized within the XPFC formulation for $\ffpfc[\phii]$, 
see \cite{GPR10,OSP13}. The correlation function $C_2$ is approximated in k-space as the
envelope of a set of Gaussians and with peaks chosen by the
primary k-vectors defining the crystal structure. For a square symmetry a
minimum of two peaks is needed, $\widehat{C}_2(\kk) =
\max{(\widehat{C}_{2,0}(\kk), \widehat{C}_{2,1}(\kk))}$ and
$\widehat{C}_{2,i}(\kk)=A_i \exp{[(\kk_i-\kk)^2/(2 \xi_i^2)]}$.  
The effect of temperature on the elastic properties is seen in
the width of the peaks and modeled by $\xi_i$.   $A_i$ is a Debye-Waller
factor controlling the height of the peaks.

Magnetization in an isotropic and homogenous material is modeled by
$\ffm[\mm]$. The last two terms describe the interaction of the
magnetization with an external and a self-induced magnetic field, $\BBext$ and
$\BBind$, respectively.  The magnetic field is defined as $\BB=\BBext+\BBind$,
where \BBind is defined with help of the vector potential: $\BBind=\nabla
\times \VA $ and $ \nabla^2 \VA = - \nabla \times \mm$. 

The magnetic anisotropy of the
material is due to the crystalline structure of the material. Thus, the
magnetization has to depend on the local structure represented by \phii and
vice versa. The first term in $\ffcoup[\phii,\mm]$, changes the ferromagnetic
transition in the magnetic free energy. On average $\phii^2$ is larger in the
crystal than in the homogeneous phase. Thus, $\omega_m$ and $r_m$ can be chosen
to realize a paramagnetic homogeneous phase and a ferromagnetic crystal. The
second term depends on average on the relative orientation of the crystalline
structure with respect to the magnetization.

The number density \phii evolves according to conserved
dynamics and magnetization according to non-conserved dynamics,
\begin{align} \label{eq1}
\frac{\partial \phii}{\partial t} = M_n \nabla^2 \frac{\delta \FF[\phii,\mm]}{\delta \phii}, \quad
\frac{\partial \mmi}{\partial t} = - M_m \frac{\delta
\FF[\phii,\mm]}{\delta \mmi}
\end{align}
$i = 1,2$, respectively. 
However, in the limit of strong external magnetic fields, \BBext, the magnetization, \mm, can assumed to be homogeneous in the crystal. 
As shown in Ref.~\cite{BEV19} the magnetization becomes perfectly aligned with the external magnetic field and independent of the relative orientation 
of the crystal. For paramagnetic or ferromagnetic materials near the Curie temperature, the magnitude of the magnetization $m = |\mm|$ 
dependents on the magnitude of the external magnetic field \BBext. In this limit $f_m(\mm)$ is constant and does not influence the dynamics. Furthermore, we are only concerned with the crystal phase and assume $\omega_m = 0$.
The remaining parameters are chosen as in Table \ref{tab:param} and lead to a minimization of energy if the magnetization is aligned with 
the \hkl<1 1>-directions of the crystal, the easy axis. The hard axis are the \hkl<1 0>-directions.
Thus, a preferably or perfectly aligned single crystal has a \hkl<1 1>-direction aligned with the external magnetic field.
Due to the direct relation between \BBext and \mm, only the evolution equation for \phii remains and reads:
\begin{align} \label{eq2}
\frac{\partial \phii}{\partial t} = M_n \nabla^2 &\left[\vphantom{\sum_{j=1}^{2}} \phii - \frac{t}{2} \phii^2 + \frac{v}{3} \phii^3 \right. \\
& - \int  C_2(\rr-\rr') \phii[\rr'] \ddr' \nonumber \\
&\left.+ \omega_B \nabla \sum_{j=1}^{2} \alpha_{2j} (\mm \nabla\phii)^{2j-1} \mm \right], \nonumber
\end{align}
where $\mm$ is considered as a parameter. Increasing \mm leads to increasing anisotropy and magnetostriction \cite{BEV19}.
The external magnetic field \BBext and thus \mm is assumed to be parallel to the thin film. Thus, in this limit of strong external magnetic fields we can use \mm to vary the strength and direction of the influence of the external magnetic field on the thin film. The magnitude of \mm is varied between [0; 0.8] and varies the magnetic anisotropy.

\begin{table}
\begin{tabular}{|c|c|c|c|c|c|c|c|c|}
\hline
t  & v & M$_{n}$ & $\bar{\phii}$ & $k_{0/1}$                                   & $\xi_{1/2}$          & $A_{0,1}$            & $\omega_B$ & $\alpha_{2,4}$ \\ \hline
1 & 1 & 1            & 0.05             & $\left( 2 \pi, \sqrt{2} 2 \pi \right)$ & $\left(1,1 \right)$ & $\left(1,1 \right)$ & 1                    & (-0.001, 0) \\ \hline
\end{tabular}
\caption{\label{tab:param} Modeling parameters. The parameters are inspired by \cite{OSP13} and chosen to maximize the energetic difference between square and triangular phase. }
\end{table}
In order to increase numerical stability, 
short wavelength in the solutions of the 
density are gradually damped in k-space by adding $-10^{-6}(2 \kk_1 - \kk)^2$ to 
C$_k$(\kk). 
The evolution equation is solved semi-implicitly in time with a pseudo-spectral method. For numerical details we refer to Refs.~\cite{Backofenetal_PML_2007,Praetoriusetal_SIAMJSC_2015}. The reduced model Eq. \eqref{eq2} is numerically more 
stable and less costly compared to the full model \eqref{eq1}. The timestep may be increased by an order of magnitude. Thus, coarsening simulations for large times become feasible. 

Here, the thin film is modeled by a two dimensional slab perpendicular to the film height. The crystalline order is defined by the density wave, \phii. The external magnetic field is assumed parallel to the film and induces a homogeneous magnetization. The magnetic driving force in the model is controlled by the magnitude of the magnetic moments.  

We choose a parameter set, which shows stagnation in coarsening to include the effect of retarding forces and reflect the experimental findings. 
The simulation domain has size $L^2=819.2^2$. The mean distance of density peaks is one and is resolved by ten grid points, (dx=0.1). Thus, the whole systems consists of $6.7\cdot10^5$ density peaks, representing particles. A time step of dt=0.1 was used.

\section{Coarsening}
Equation \eqref{eq2}, is used to model magnetic assisted annealing of thin films. The texture of the polycrystalline 
structure is monitored during annealing in order to extract geometrical and topological properties over time 
and compare them for different magnitudes $m$.
To generate an appropriate initial condition we set \mm = 0, start with a
randomly perturbed density field \phii, and solve eq. \eqref{eq2} until we reach a polycrystalline structure with small crystallites with square symmetry. The perturbation is a random distortion at every grid point. The small wavelength perturbations are smoothed rapidly by the evolution equation, but long wavelength perturbation act as nucleation centers. Thus, at random positions grains with random orientation begin to grow until they touch and from a network of grain boundaries.  After impingement we got about 1,600 randomly oriented grains.
This configuration is used as initial condition for all simulations.

\subsection{Scaling}

\begin{figure}[h]
\raisebox{-.5\height}{\includegraphics[width=0.45\textwidth]{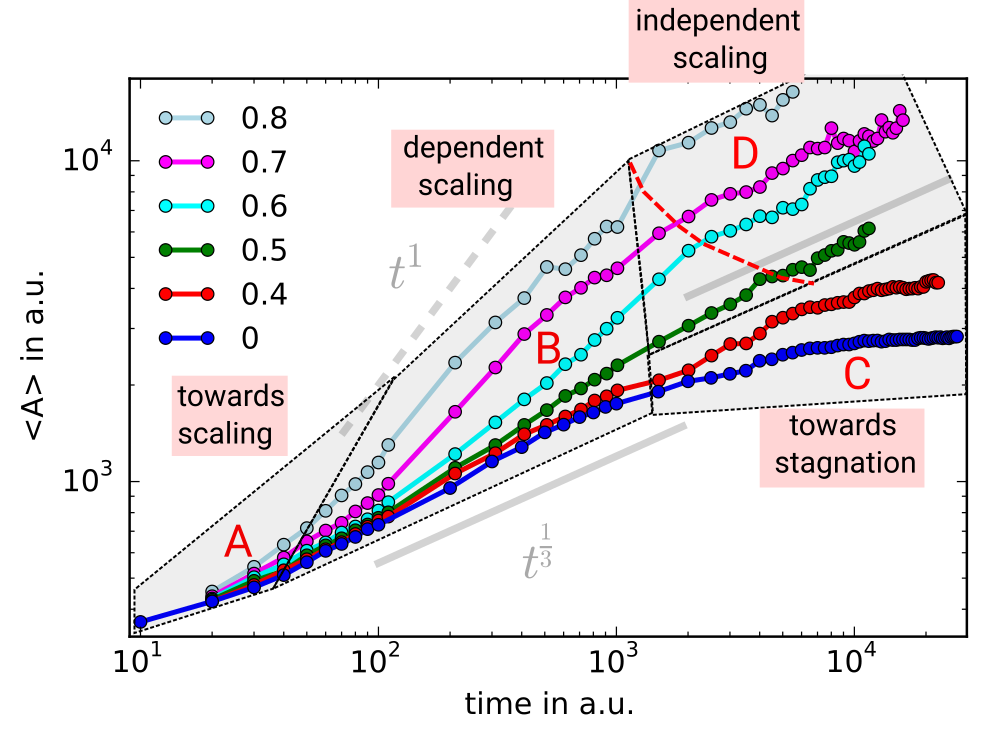}}
\caption{\label{fig:scaling}
Long time evolution of mean grain area for different magnetization. Four different regimes are identified: (A) towards scaling, an initial phase; (B) dependent scaling, a magnetically enhanced scaling regime with the scaling exponent depending on $m$; (C) towards stagnation, a regime which is only present without or with low magnetic fields; and (D) independent scaling, a regime reached at late times, with a scaling exponent independent of magnetic anisotropy. $m$ is varied between [0; 0.8] and models the strength of magnetic influence and anisotropy.}
\end{figure}

Figure~\ref{fig:scaling} shows the evolution of the mean grain area, \hkl<A>. Coarsening leads to an increase of the mean grain area over time. The coarsening is enhanced by increasing the magnetization and, thus, the magnetic anisotropy. We identify scaling regimes by a power law, $ \hkl<A>\propto t^\alpha$, with a scaling exponent $\alpha$.   
In all cases a first scaling regime Fig.~\ref{fig:scaling}(B) is reached after an initial phase Fig.~\ref{fig:scaling}(A).
Without magnetization a scaling exponent of $\alpha=1/3$ is observed. Increasing the magnetic influence increases the scaling exponent. 
The maximum scaling exponent $\alpha=1 $ is achieved for $m=0.8$. However, this scaling regime ends. For small magnetic influence below some threshold, it turns into stagnation, Fig.~\ref{fig:scaling}(C).  
Above this threshold, here $m \ge 0.5$, the scaling becomes independent of magnetic interaction and we observe $\alpha=1/3$, 
Fig.~\ref{fig:scaling}(D). 

It has been shown before that without magnetic driving force the scaling exponent depend on initial conditions and modeling parameters \cite{BBV14}. This also remains if magnetic driving forces are included. The identified regimes (A), (B), (C) and (D) thus also depend on initial conditions and modeling parameters.

Without magnetic driving force the texture becomes self similar during coarsening \cite{Barmaketal_PMS_2013,BBV14}. This is not the case for magnetically enhanced coarsening due to grain selection. In the following we analyze texture evolution during coarsening in detail in order to understand the change of the scaling behavior. 


\subsection{Orientation selection}

\begin{figure*}
      \hspace*{-1cm}\raisebox{-.5\height}{ \includegraphics[width=0.85\textwidth]{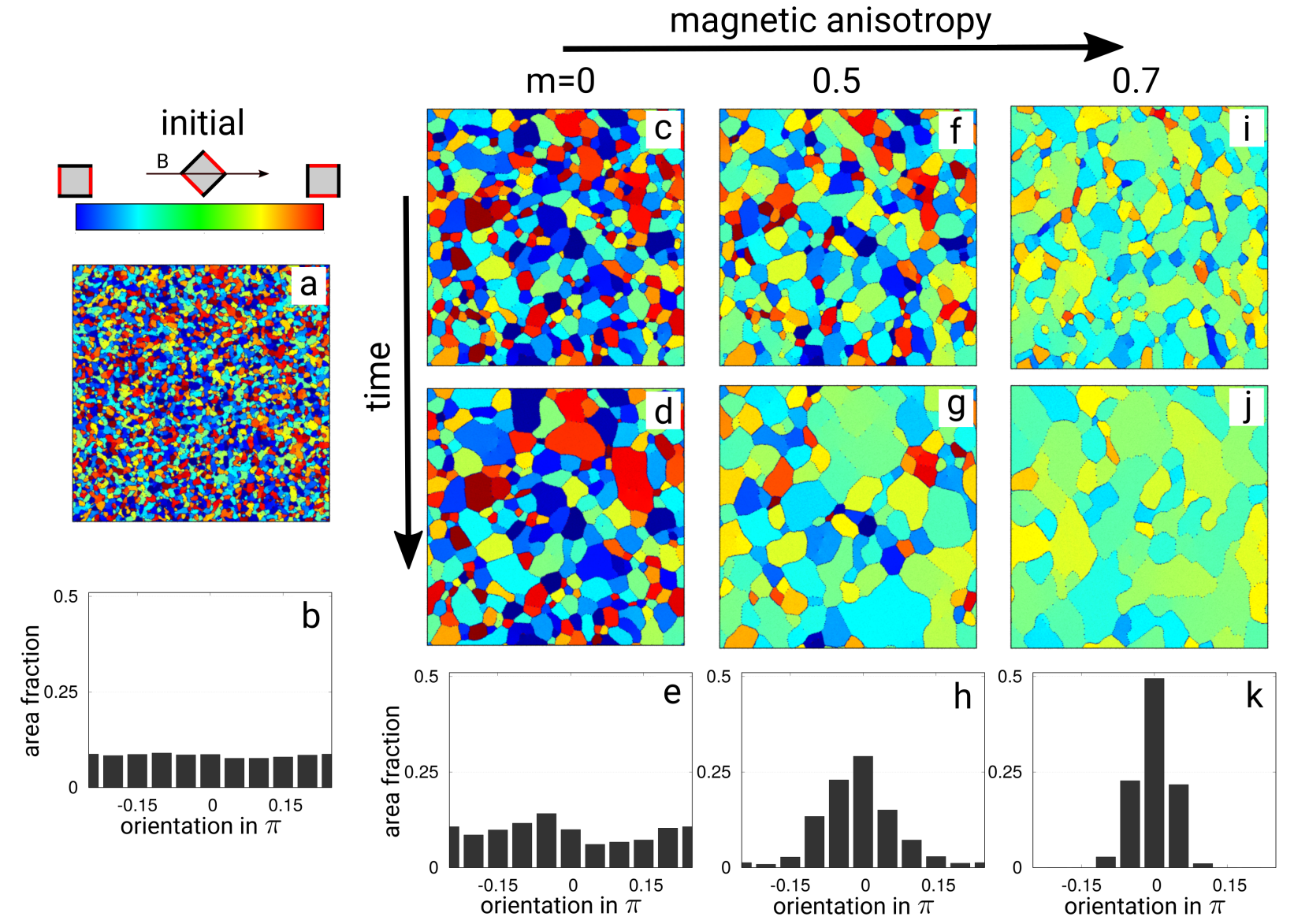}} 
      \caption{\label{fig:orient}
        Grain structure during annealing. The color represents the local orientation of the easy axis with respect to the external magnetic field. The area fraction is shown as function of orientation for the initial and final configurations for different magnetic fields $m$. The times for the snapshots for m=0, 0.5 and 0.7 are ($9\cdot10^3, 2.7\cdot10^4$), ($1\cdot10^3, 1.1\cdot10^4$) and  ($4.1\cdot10^3, 1.6\cdot10^4$), respectively.}
\end{figure*}

The magnetic driving force leads to preferable growth of grains, which are preferably aligned with respect to the external magnetic field. Figure~\ref{fig:orient} shows typical orientation distributions and how they evolve over time dependent on the magnetic influence. The color represents the local crystal orientation, $\theta$. A preferably aligned crystal corresponds to $\theta =0$ and, due to symmetry, the $\theta$ varies in the range $[-0.25 \pi, 0.25 \pi]$.

The initial orientation distribution is constructed without magnetization. Thus, it is homogeneous, Figs.~\ref{fig:orient}(a) and (b). There is no preferred orientation for the grains. Without a magnetic driving force, $m=0$, the orientation distribution stays homogeneous, Figs.~\ref{fig:orient}(c)-\ref{fig:orient}(e). With a magnetic driving force this changes and well aligned grains grow preferentially, Figs.~\ref{fig:orient}(f)-~\ref{fig:orient}(k). Grains with $\theta\approx 0$ (green) grow at the expense of the other grains (blue, red). As already quantified in Fig.~\ref{fig:scaling}, the enhanced grain growth with increasing $m$ can be seen also by larger grain sizes for increasing $m$, Fig.~\ref{fig:orient}(d),~\ref{fig:orient}(g) and ~\ref{fig:orient}(j). However, we are here interested in the orientation distribution, which becomes sharply peaked at $\theta = 0$, Fig.~\ref{fig:orient}(e),~\ref{fig:orient}(h) and \ref{fig:orient}(k). The effect increases with increasing magnetic driving force, as already analyzed for the full model \eqref{eq1} in Ref.~\cite{BEV19}.  

\begin{figure}[h]
  \begin{tabular}{c}
    \raisebox{-.5\height}{ \includegraphics[width=0.25\textwidth]{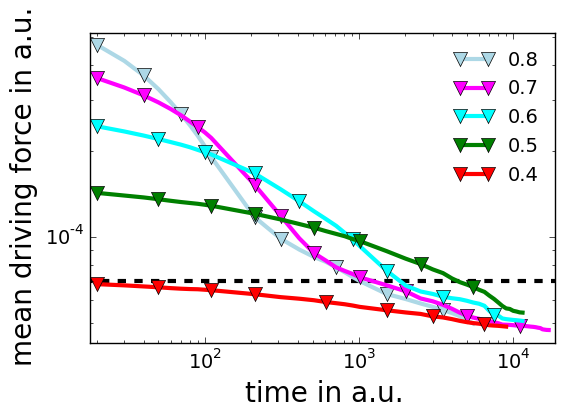}} 
\end{tabular}
\caption{\label{fig:mmdf} Mean magnetic force during coarsening for different applied magnetic fields $m$. }
\end{figure}

The narrowing in orientation distribution has an effect on the total impact of the external magnetic field. As it reduces the mean orientation difference of adjacent grains it also reduces the mean magnetic driving force. To measure this effect we define the mean magnetic driving force as the average energy difference due to magnetic anisotropy with respect to a perfectly aligned crystal. Figure~\ref{fig:mmdf} shows this quantity over time. Initially the mean magnetic driving force strongly depends on the strength of the magnetic field. Large $m$ lead to large magnetic anisotropy and, thus, large magnetic driving forces. However, over time the mean magnetic driving force decreases as the mean orientation deviation from a perfectly aligned crystal decreases due to grain selection. The strength of this effect correlates with the strength of the magnetic field. At large times, the mean magnetic driving force falls below a threshold. This large time behavior correlates with the independent scaling regime in Fig.~\ref{fig:scaling}(D), which occurs, when the mean magnetic driving force falls below $\approx 0.7 \cdot 10^{-4}$. The time this threshold is reached depends on $m$ and is indicated by the dashed (red) line in Fig.~\ref{fig:scaling}. Thus, orientation selection induced by the external magnetic field over time decreases the influence of the magnetic field, which explains the transition to the independent linear scaling (D) in Fig.~\ref{fig:scaling}. 
 In the case of stagnation, $m<0.5$, the mean magnetic driving force never exceeds the defined threshold.  

\subsection{Grain Size Distribution}
\begin{figure*}
  \begin{tabular}{ccc}
   \raisebox{-.5\height}{ a)\includegraphics[width=0.3\textwidth]{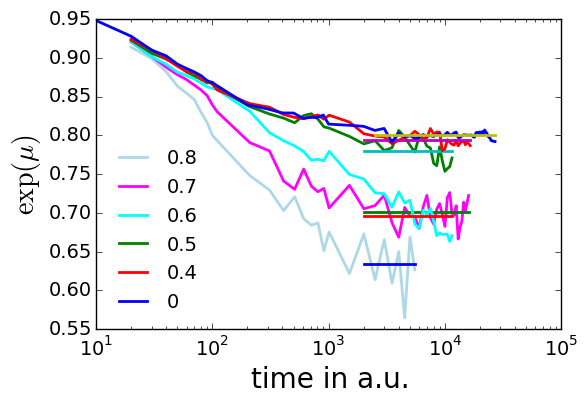}} &
    \raisebox{-.5\height}{ b)\includegraphics[width=0.3\textwidth]{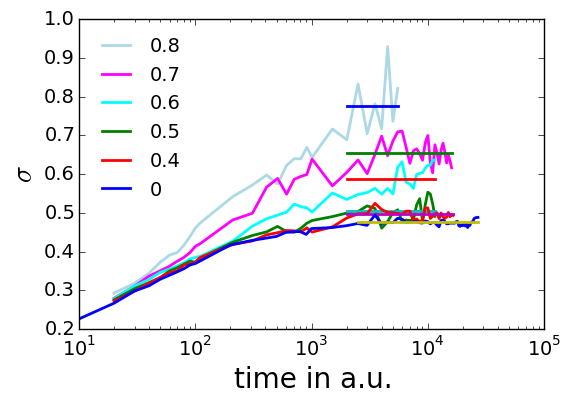}} &
    \raisebox{-.5\height}{ c)\includegraphics[width=0.3\textwidth]{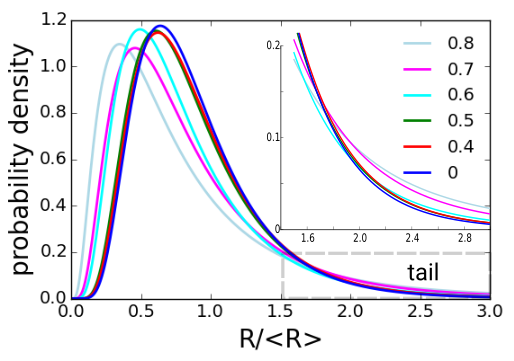}}    
\end{tabular}
\caption{\label{fig:GSD} Log-normal distribution parameters $\exp(\mu)$ (a) and $\sigma$ (b) over time and GSD (c) for final averaged values for $m$ between [0; 0.8]. The data for $m=0$ correspond with \cite{BBV14} and the experimentally found universal GSD in \cite{Barmaketal_PMS_2013}.}
\end{figure*}

The external magnetic field does not only change the orientation distribution but also the grain size distribution (GSD). Without external magnetic fields it was shown in \cite{BBV14} that the coarsening becomes self similar and the GSD is well described by a log-normal distribution: $(\sqrt{2 \pi} \sigma x)^{-1} \exp{(-\frac{(\log{x}-\mu)^2}{2 \sigma^{2}})}$, where x is the scaled radius $\frac{R}{\hkl<R>}$.
We calculate the GSD for all coarsening simulations and fit log-normal distributions to our results. In Figs.~\ref{fig:GSD}(a) and \ref{fig:GSD}(b) the two values defining the log-normal distribution, $\exp(\mu)$ and $\sigma$ are shown over time. During the dependent (magnetically enhanced) scaling, Fig.~\ref{fig:scaling}(B), $\exp(\mu)$ and $\sigma$ change: $\exp(\mu)$ decreases, while $\sigma$ increases. Thus, the GSD is not constant over time and, thus, the coarsening is not self similar. Only within the independent scaling regime and towards stagnation, Figs.~\ref{fig:scaling}(C) and \ref{fig:scaling}(D), the GSD becomes stationary on average. Thus, self similar growth is achieved. 

As the number of grains is drastically decreased within this regime the GSD statistics become more and more noisy. Fluctuations in the GSD approximation increase for larger times and higher magnetic influence. In order to compare the GSD for different external magnetic fields in the limit of large times, we average $\exp(\mu)$ and $\sigma$ for large times and use the averaged value to reconstruct the log-normal distribution, see Fig.~\ref{fig:GSD}(c). Large external magnetic fields, $m > 0.5$ shift the maximum of the GSD towards smaller sizes. However, the tail becomes wider. Thus, the number of large grains with respect to the average grain size is increased. For smaller external magnetic fields, $m < 0.5$ the tendency is the same but the difference is minor.            

\subsection{Grain coordination and shape}

Various other geometrical and topological measures have been considered to define the grain structure.
The next neighbor distribution (NND) or coordination number of grains counts the number of neighboring grains. The shape of grains can be quantified by approximating every grain by an ellipse. The ratio of the axis of the ellipse then measure the elongation of grains. This leads to the axis ratio distribution (ARD). Elongated grains my have preferred direction of elongation. This is measured here by the angle of the small axis with the external magnetic field and lead to a small axis orientation distribution (SAOD).    
   
\begin{figure*}
  \begin{tabular}{ccc}
    \raisebox{-.5\height}{ a)\includegraphics[width=0.3\textwidth]{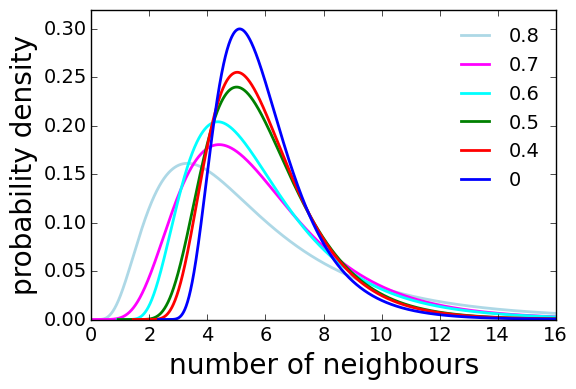}} &
    \raisebox{-.5\height}{ b)\includegraphics[width=0.3\textwidth]{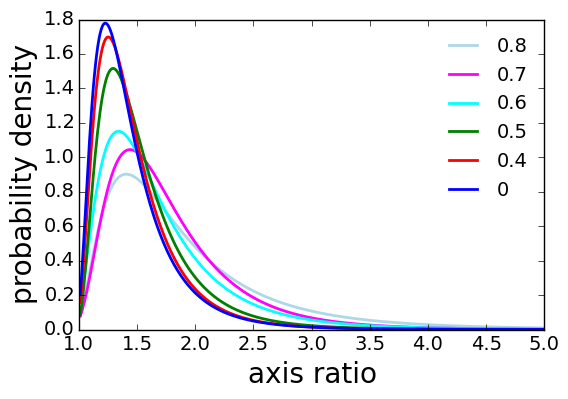}} &
    \raisebox{-.5\height}{ c)\includegraphics[width=0.3\textwidth]{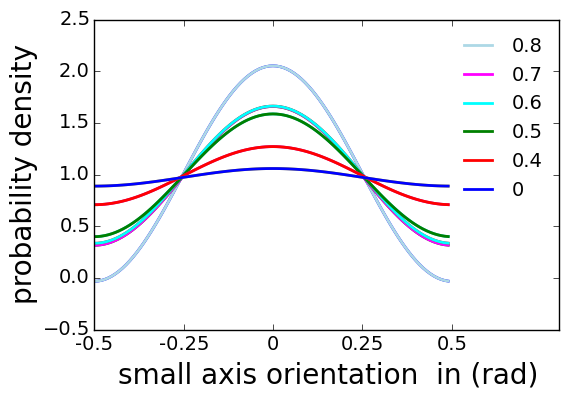}}
\end{tabular}
\caption{\label{fig:XXD}  Log-normal description for next neighbor distribution (NND) (a), the smoothed distribution should of course be interpreted in a discrete setting, axis ratio distribution (ARD) (b) and cosine description for small axis orientation distribution (SAOD) (c), obtained from late time coarsening regime. $m$ is varied between [0; 0.8]. Note: the NND has only discrete values and isrepresented by a smooth density distribution to show the influence on m.}
\end{figure*}

We concentrate on large times for which the coarsening is self similar. Figure~\ref{fig:XXD}(a) shows the NND, which is also fitted by a log-normal distribution. With increasing external magnetic field the distribution broadens and the maximum is shifted to smaller values. This can already be related to the faster growth, which leads to larger grains and thus also an increased difference in grain size. Classical empirical laws for topological properties in grain structures, such as the Lewis' law and the Aboav-Weair's law, see Ref.~\cite{Chiu_MC_1995} for a review, show a linear relation between the coordination number and the area of the grains and postulate that grains with high (low) coordination number are surrounded by small (large) grains, respectively. These effects are further enhanced by the elongation of grains, which lead to more neighbors. Additionally, small grains between elongated grains have less neighbors. 

The ARD can also be approximated by a log-normal distribution, Fig.~\ref{fig:XXD}(b). With increasing magnetic anisotropy the ratio increases and more and more elongated grains are present. The orientation of the elongation is correlated with the external magnetic field. In Fig.~\ref{fig:XXD}(c) the orientation distribution of the small axes with the direction of the external magnetic field (SAOD) is shown. Here the distribution is approximated by a cosine. The elongated grains become more and more oriented perpendicular to the external magnetic field.  


\section{Discussion}
Classical Mullins-like models for grain growth predict self similar growth and a scaling law $\hkl<A> \propto t^\alpha$ with a scaling exponent $\alpha = 1$ \cite{Mul98}. This also does not change if external magnetic fields are introduced as an additional driving force. In contrast to our simulation, see Fig. \ref{fig:scaling}, no influence of the scaling behavior is observed. Even though the texture depends on strength and direction of the external magnetic field \cite{MKM06,BMM07,MBM07,LZS09,All16,GMT18}. In these simulations the increase of growth of well aligned grains is leveled by the decrease of growth of not well aligned grains. Thus, the scaling exponent is predicted to be independent of the additional driving force. In these models, smaller exponents and stagnation of grain growth, as observed in experiments \cite{Barmaketal_PMS_2013}, can only be achieved by introducing additional retarding or pinning forces. 

Within the considered PFC model triple point and orientational pinning are naturally present, which is one reason for the observed lower scaling exponent and the stagnation~\cite{BBV14}. External magnetic fields introduce an additional driving force to the system. If large enough they can overcome the retarding forces and enhance growth. This explains the dependent growth regime with scaling exponents depending on the applied magnetic field. If the magnetic driving force is large enough all retarding forces are overcome and an exponent of $\alpha = 1$ is reached.

Grain growth under an applied magnetic field leads to preferable growth of well aligned grains. It is this grain selection which decreased the mean magnetic driving force over time. If the texture is dominated by well aligned grains, the magnetic driving force is no longer a function of the applied field but is limited by the texture, see Fig. \ref{fig:mmdf}. Only parts of the retarding forces can be overcome and the scaling exponent becomes independent of the magnetic interaction. Turning off the magnetic field in this regime of well aligned grains leads to stagnation. It can only be speculated about the origin of this retarding forces and the mechanism they are overcome by the magnetic field. However, crystalline defects and elastic properties are known to be modified by the local magnetization \cite{BEV19} and lead to magnetization dependent mobilities. The same mechanism may also open new reaction paths for defect movement which might remove the retarding forces. 

In the case of small magnetic field the coarsening stagnates. In this regime the magnetic driving force is not large enough to overcome the retarding force responsible for stagnation.

Within the independent scaling regime self similar growth is observed which allows to compute various geometrical and topological properties of the grain structure. Their dependence on the magnitude of the applied magnetic field has been analyzed. The considered grain size distribution (GSD), next neighbor distribution (NND) and axis ratio distribution (ARD) broaden with increasing magnetic anisotropy, leading to larger grains, more grains with very few and many neighbors, and more elongated grains, see Figs. \ref{fig:GSD} and \ref{fig:XXD}. The shift in the NND to smaller coordination number has also been reported for simulations based on Mullins type models \cite{MBM07}.

\begin{figure}
      \hspace*{-1cm}\raisebox{-.5\height}{ \includegraphics[width=0.3\textwidth]{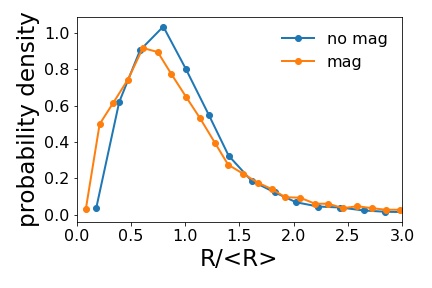}} 
      \caption{\label{fig:exp}
        GSD of Zr sheet after annealing with and without magnetic fields of 19 T. Data is extracted from Fig.~14 in Ref.~\cite{MB10}. The mean grain size \hkl<R> is $10 \mu\rm{m}$ for the sample annealed without magnetic field and  $18 \mu\rm{m}$ for the sample annealed with magnetic field. 
}
\end{figure}
Even though, texture control by magnetic fields is of increasing interest~\cite{Riv13} there are not much data on the influence of magnetic fields on GSD in thin films available. 
In \cite{MB10} the texture and grain size evolution of thin Zr sheets annealed with and without magnetic fields at different temperatures are studied. Increasing temperature and applying external magnetic fields lead to increasing mean size of the grains. The orientation of the final grains are influenced by the magnetic field and the orientation distribution becomes peaked at favorable orientations. The same tendency as predicted by our simulations, Fig.~\ref{fig:orient}. In Fig.~\ref{fig:exp} the GSD are compared for these samples after annealing with and without magnetic field. The magnetic field shifts the peak of the GSD towards smaller values leading to an increase of relatively small grains and relatively large grains. The GSD also widens and the tail is increased by the magnetic field. Also these details in the evolution follow qualitatively our simulation results, Fig.~\ref{fig:GSD}.
But we are not aware of an experimental study showing the increased elongation of the grains perpendicular to the external magnetic field.      \\

\section{Conclusion}

We studied magnetically enhanced coarsening with an extended PFC model. The external magnetic field is assumed to be strong enough to prescribe the magnetization of the thin film. That is, the magnetization is constant and perfectly aligned with the external magnetic field. The anisotropy of the magnetic properties of the crystal lead to a magnetic driving force. Well aligned crystals grow at the expense of not well aligned crystals. Additionally, magnetostriction leads to deformation of crystal and defect structures.

The magnetic driving force leads to grain selection and a texture dominated by well aligned grains. As the amount of similar oriented grains increase, the mean orientation difference between grains decreases. Thus, the mean magnetic driving force also decreases with time due to texture change. The scaling exponent becomes independent for large times and for large enough magnetization. Stagnation and variation of scaling exponents is due to retarding and pinning forces for grain boundary movement.  
There are two mechanisms in magnetically enhanced coarsening, which change the effect of retarding forces. Firstly, the magnetic driving force helps to overcome the retarding forces during coarsening. This explains the scaling regime dependent on the magnetic anisotropy.
Secondly, the change of structure of the crystal due to magnetostriction can decrease the energy barriers representing the retarding force. Then the driving force due to minimization of grain boundary energy may become large enough to overcome the retarding forces. This could explain the independent scaling regime.   

But not only the scaling changes, characteristic geometric and topological properties are also influenced by the applied magnetic field. At least for GSD and NND experiments show the same tendency as predicted by our simulations. 

\begin{acknowledgments}
A.V. and R.B. acknowledge support by the German Research Foundation (DFG) under Grant No. Vo899/21 in SPP1959.
We further gratefully acknowledge the Gauss Centre for Supercomputing e.V. (www.gauss-centre.eu) for funding this project (HDR06) by providing computing time through the John von Neumann Institute for Computing (NIC) on the GCS Supercomputer JUWELS at J\"ulich Supercomputing Centre (JSC).
\end{acknowledgments}

\bibliography{MagneticCoarsening}
\newpage
\appendix


\end{document}